\documentclass[a4paper,10pt]{article}

\usepackage{amsmath}
\usepackage{amssymb}
\usepackage{amsthm}
\usepackage{graphicx}

\newtheorem{definition}{Definition}
\newtheorem{theorem}{Theorem}
\newtheorem{proposition}{Proposition}
\newtheorem{lemma}{Lemma}
\newtheorem{corollary}{Corollary}
\newtheorem{remark}{Remark}

\renewcommand{\epsilon}{\varepsilon}
\renewcommand{\phi}{\varphi}

 \DeclareMathOperator{\trace}{Tr}
 \DeclareMathOperator{\I}{I}
\DeclareMathOperator{\littleo}{o}

\renewcommand{\Re}{\mathop{\rm{Re}}\nolimits}
\renewcommand{\Im}{\mathop{\rm{Im}}\nolimits}

\newcommand{\norm}[1]{\left\Vert #1\right\Vert}
\newcommand{\module}[1]{\left| #1\right|}

\newcommand{\R}{\mathbb{R}}
\newcommand{\C}{\mathbb{C}}
\newcommand{\E}{\mathbb{E}}

\newcommand{\indicator}[1]{\mathbf{1}_{#1}}

\newcommand{\tens}{\otimes}

\newcommand{\scal}[2]{\langle #1 , #2\rangle}

\newcommand{\ketbra}[2]{| #1 \rangle \langle #2 |}

\newcommand{\ket}[1]{| #1 \rangle}

\newcommand{\mcH}{\mathcal{H}}
\newcommand{\mcK}{\mathcal{K}}
\newcommand{\mcE}{\mathcal{E}}
\newcommand{\mcU}{\mathcal{U}}
\newcommand{\mcD}{\mathcal{D}}
\newcommand{\mcM}{\mathcal{M}}

\title{Asymptotics of random density matrices}
\author{Ion Nechita\footnote{Universit\'e de Lyon, Universit\'e Lyon 1, CNRS, UMR 5208 Institut Camille Jordan, Batiment du Doyen Jean Braconnier, 43, blvd du 11 novembre 1918, F - 69622 Villeurbanne Cedex, France. E-mail: \texttt{nechita@math.univ-lyon1.fr}}}
\date{}

\begin{document}
\maketitle
\begin{abstract}
We investigate random density matrices obtained by partial tracing larger random pure states. We show that there is a strong connection between these random density matrices and the Wishart ensemble of random matrix theory. We provide asymptotic results on the behavior of the eigenvalues of random density matrices, including convergence of the empirical spectral measure. We also study the largest eigenvalue (almost sure convergence and fluctuations).
\end{abstract}

\section{Introduction}

Physicists and computer scientists working with finite size quantum systems are often interested in properties of \emph{typical} states, such as entanglement, entropy, etc. In order to estimate such quantities, one has to endow the set of states (pure or mixed) with a certain probability measure and compute averages with respect to this measure. It has been known for a certain while now that there exists an "uniform" (in a way which will precised later) measure on the set $\mcE_n$ of pure states of size $n$. However, the situation is less simple when dealing with density matrices: there is no widely accepted candidate for a "canonical" measure on the set $\mcD_n$ of all density matrices of size $n$. 

One may find in the literature two classes of probabilities on $\mcD_n$:
\begin{itemize}
\item the \emph{induced measures}, where random density matrices are obtained by partial tracing a larger random pure state,
\item the \emph{metric measures}, where the measure is the volume element associated to a particular distance on $\mcD_n$.
\end{itemize}

Depending on the physical nature of the model, one may choose different measures from one class or the other. In this work we study the measures of the first class. 

The induced measures were introduced by Braunstein \cite{braunstein} and studied later by Hall \cite{hall}, {\.Z}yczkowski and Sommers \cite{sz1, sz2}. In the first part of this work we provide a rigorous construction of these measures. In the second part, we give some new explicit and recurrence formulas for the moments and we study the asymptotic behavior of the spectrum of such random density matrices. Our approach is based on the connection with the well-known theory of \emph{Wishart} random matrices.

Our paper is organized as follows. In section \ref{sec:pure_states_and_density_matrices} we recall the construction of the induced measures, adding mathematical rigor to the existing literature. Section \ref{sec:wishart_fixed_size} is devoted to recalling some results on the Wishart ensemble and making explicit the connection with random density matrices. We deduce the distribution of the eigenvalues and we study the moments. In Section \ref{sec:asymptotics} we study two models of large random density matrices, providing results on the behavior of the spectrum. A discussion of the results as well as ideas for generalizing our work are presented at the end of the paper.

\section{From pure states to density matrices}\label{sec:pure_states_and_density_matrices}
We start by introducing and motivating the model of random density matrices that we consider.

As explained in the Introduction, one would like to endow the set of density matrices on a complex Hilbert space $\mcH$ with a ``natural'' probability measure. It turns out that there is no straightforward way of doing this, so one has to make some additional hypothesis in order for a probability measure to stand out as the most natural one. Our approach here is based on the definition of a density matrix as it is usually understood in the theory of open quantum systems. We consider that the system described by the density matrix is coupled to an environment and that the compound system is in a random pure state. More precisely, we shall make two assumptions:
\begin{enumerate}
\item[(A1)] The system $\mcH$ is \emph{coupled} to an environment $\mcK$ and the compound system $\mcH \tens \mcK$ is in a pure state $\ket \psi$.
\item[(A2)] The pure state $\ket \psi$ is \emph{uniformly} distributed on the set of all pure states on $\mcH \tens \mcK$.
\end{enumerate}

The first assumption is motivated by a large class of models considered in physics or quantum information theory. The general framework is provided by a system $\mcH \tens \mcK$ in a pure state, isolated from its environment. Suppose that one has access only to the sub-system $\mcH$. This may happen for several different reasons: $\mcK$ may be not accessible (e.g. $\mcH$ and $\mcK$ are in distant galaxies) or $\mcK$ may be too complicated to study (a heat bath or a noisy channel, for example). In these situations, it is natural to make the assumption (A1). Let us turn now to the second assumption. If one has no \emph{a priori} information on the systems $\mcH$ and $\mcK$, it makes sense to suppose (A2). Moreover, it turns out that there exists an unique uniform probability measure on the set of pure states of given size, so we shall consider \emph{uniform} pure states on the compound system. 

However, there are situations when one of the two hypotheses (A1) or (A2) is not physically motivated. For instance, when one has no knowledge of an environment coupled to the system $\mcK$, there is no reason to suppose (A1). Instead, one should use other probability measures, such as the \emph{Bures measure} (see the discussion in \cite{sz1}). On the other hand, even if (A1) corresponds to the physical reality, one may have extra information on the system $\mcH$ or $\mcK$ (or both). For example, it may be that the state of the environment $\mcK$ has a particular form; thus, it makes no sense to assume (A2) and our model would not be adapted to such situations.

In the next section, motivated by the assumption (A2), we shall construct the uniform measure on the set of pure states of given size. Then, by partial tracing, we shall provide the probability which verifies the assumptions (A1) and (A2).

\subsection{The canonical probability measure on the pure states}

In quantum mechanics, a pure state is described by a norm one vector in a $n$-dimensional complex vector space $\mcH$. The phase of pure states is not determined, i.e. 
\begin{equation}
\ket{e^{i\theta} \psi} = \ket \psi \quad \forall \,\theta \in \R
\end{equation}

In order to make this definition rigorous, we introduce the following equivalence relation on $\mcH \smallsetminus \{0\}$:
\begin{equation}
x \sim y \Leftrightarrow \exists \,\lambda \in \C^* \textit{ such that } x =\lambda y.
\end{equation}

\begin{definition}
A pure state $\ket \psi$ is an element of the quotient space $(\mcH \smallsetminus \{0\}) / \sim$. We denote by $\mcE_n$ the set of pure states of size $n$.
\end{definition}

As all complex Hilbert spaces are isomorphic to $\C^n$, the set $\mcE_n$ is the set of rays in $\C^n$. We endow $\mcE_n$ with the associated quotient topology and the Borel $\sigma$-field. We now turn to the construction of the uniform probability measure on $\mcE_n$.

As stated in the assumption (A2), the probability on $\mcE_n$ should be the most \emph{uniform} one, as there is no \emph{a priori} information on the state $\ket \psi$. In particular, as there is no preferred basis of $\mcH$, the uniform measure should be invariant by changes of bases. In our framework ($\mcH$ is a complex Hilbert space), changes of bases are provided by unitary applications. As a consequence, we ask that the uniform probability measure should be unitarily invariant.
\begin{definition}
A measure $\nu_n$ on $\mathcal E_n$ is \emph{unitarily invariant} if
\begin{equation*}
\nu_n(UA) = \nu_n(A),
\end{equation*}
for all unitary $U \in \mcU(n)$ and for all Borel subset $A \subset \mathcal
E_n$.
\end{definition}

It turns out that this condition is strong enough to completely specify a measure on $\mcE_n$, i.e. there is an unique unitarily invariant probability measure on $\mcE_n$. This follows from a well-known result in probability theory regarding group actions (see \cite{kallenberg}). Let us recall it here.

Let $G$ be a topological group acting on a topological space $X$. We call its action \emph{transitive} if for all $x, y \in X$, there exists $g \in G$ such that $y = g\cdot x$ and \emph{proper} if for all $g \in G$, the application $X \ni x \mapsto g \cdot x$ is proper, i.e. the pre-image of a compact set is compact. We then have the following
\begin{theorem}[\cite{kallenberg}]\label{th:invariant}
Let $G$ be a topological group which acts transitively and properly
on a topological space $X$. Suppose that both $G$ and $X$ are
locally compact and separable. Then there exists an unique (up to a
constant) measure $\nu$ on $X$ which is $G$-invariant.
\end{theorem}

In order apply this result to our situation, we consider the action of the unitary group $\mcU(n)$ on the set $\mcE_n$ by left multiplication. We obtain the following proposition.
\begin{proposition}
The action of $\mathcal U(n)$ on $\mathcal E_n$ is transitive and
proper and thus there exists an \emph{unique} unitarily invariant
probability measure $\nu_n$ on $\mathcal E_n$.
\end{proposition}
\begin{proof}
First of all, notice that this action is well defined: the class $\ket{U\psi}$ does not depend on $\psi$, but only on the class $\ket \psi$; we say that the multiplication by an unitary is a \emph{class application}. In order to show that the action is transitive, consider two classes $\ket \psi$ and $\ket \phi$ and an unitary $U \in \mcU(n)$ such that $U\psi = \phi$ (such an unitary always exists). It follows then that $U \ket \psi = \ket \phi$. Finally, the action is compact, as the set $\mcE_n$ is compact and the multiplication applications are continuous. Thus, the action verifies the hypothesis of Theorem \ref{th:invariant}, and as a consequence there is an unique unitarily invariant measure on $\mcE_n$. Moreover, given the compacity of $\mcE_n$ we can choose the measure of unit mass, which concludes the proof of the Proposition.
\end{proof}
Existence and unicity being settled, one would like to dispose of
more concrete descriptions on the distribution $\nu_n$. It turns out
that there are two ways of doing that.

First of all, let us recall the definition of a complex Gaussian random variable. Let $X$ and $Y$ be two independent real Gaussian random variables of mean $0$ and variance $1/2$. Then $Z = X + iY$ is said to have a complex Gaussian distribution of mean $0$ and variance $1$. We denote by $\mathcal N_\C(0,1)$ the law of $Z$. A complex vector $(Z_1, \ldots, Z_n)$ is said to have distribution $\mathcal N_\C^n(0,\I_n)$ if the random variables $Z_1, \ldots, Z_n$ are independent and have distribution $\mathcal N_\C(0,1)$.

Consider now a complex Gaussian vector $X \sim \mathcal N_\C^n(0,\I_n)$ and the projection application
\begin{align}
\Pi : \C^n \approx \mcH &\rightarrow \mcE_n\\
x &\mapsto \ket x
\end{align}
It is well-known in probability theory that the law of $X$ is unitarily invariant in $\C^n$. This property remains valid for the projection $\Pi(X)$ and thus the law of $\ket X$ is unitarily invariant on $\mcE_n$. As there is only one unitarily invariant distribution on $\mcE_n$, we have $\ket X \sim \nu_n$.

We can also obtain the law $\nu_n$ from another well-known probability distribution, the \emph{Haar measure} on $\mcU(n)$. In order to do this, consider a Haar-distributed unitary matrix $U$. Obviously, the distribution of $U$ is unitarily invariant; the same will hold true for the first column $Y$ of $U$ and  for its class $\ket Y$. Thus $\ket Y$ has distribution $\nu_n$. We sum up these results in the following
\begin{proposition}\label{prop:gaussian}
\begin{enumerate}
\item Let $X$ be a random complex vector of law $\mathcal N_\C^n(0,\I_n)$. Then the class $\ket X$ of $X$ is distributed along $\nu_n$.
\item Let $U$ be a random unitary matrix distributed along the Haar measure on $\mathcal U(n)$ and let $Y$ be the first column of $\,U$. Then the class $\ket Y$ has distribution $\nu_n$.
\end{enumerate}
\end{proposition}

\subsection{The induced measure on density matrices}\label{sec:the_induced_measures}

In this section we effectively construct the \emph{induced measures} on density matrices that will be studied in the rest of the article. As stated in the Introduction, the induced measure of parameters $n$ and $k$ is obtained as follows:
\begin{itemize}
\item Consider a product space $\mcH \tens \mcK$ of two complex Hilbert spaces $\mcH$ (of dimension $n$) and $\mcK$ - the environment - of dimension $k$. This is the global space \emph{system + environment}.
\item Take an uniform random pure state $\ket \psi$ on $\mcH \tens \mcK$ (see the assumption (A2)).
\item Consider the (random) pure density matrix $\ketbra{\psi}{\psi}$ corresponding to the pure state $\ket \psi$. 
\item Take $\rho = \trace_\mcK(\ketbra{\psi}{\psi})$, the partial trace of $\ketbra{\psi}{\psi}$ with respect to the environment $\mcK$. The law of the random variable $\rho$ is the desired probability measure, which we shall note $\mu_{n, k}$.
\end{itemize}

As in our formalism $\ket \psi$ is an equivalence class, we shall define the pure density matrix $\ketbra{\psi}{\psi}$ by:
\begin{equation}
\ketbra{\psi}{\psi} = \frac{\psi \cdot \psi^*}{\trace(\psi \cdot \psi^*)} \in \mcM_{nk}(\C).
\end{equation}

Clearly, $\psi \mapsto \ketbra{\psi}{\psi}$ is a class function (it does not depend on the representant $\psi$ chosen, but only on the class $\ket \psi$), so $\ketbra{\psi}{\psi}$ is well-defined. The normalizing factor $\trace(\psi \cdot \psi^*)$ appears because we want the matrix $\ketbra{\psi}{\psi}$ to be trace one; this could have been avoided by considering a norm one vector $\psi$, since $\trace(\psi \cdot \psi^*) = \norm{\psi}^2$.

We now turn to the third step of the above construction and recall that the partial trace is the unique application $\trace_\mcK : \mcM_{nk}(\C) \rightarrow \mcM_n(\C)$ such that
\begin{equation}
\trace((A\tens \I_\mcK)B) = \trace(A\trace_\mcK(B)) \quad \forall \, A \in \mcM_n(\C), B \in \mcM_{nk}(\C).
\end{equation}
Its expression for elementary matrices ($a_1, a_2 \in \mcH, b_1, b_2 \in \mcK$) is
\begin{equation}
\trace_\mcK[(a_1 \tens b_1)\cdot (a_2 \tens b_2)^*] = \scal{b_2}{b_1}\cdot a_1 a_2^*.
\end{equation}

We have now all the elements needed for the definition of the induced measures:
\begin{definition}
The induced measure of parameters $n$ and $k$ is the distribution $\mu_{n,k}$ of the random density matrix
\begin{equation}
\rho = \trace_\mcK(\ketbra{\psi}{\psi}),
\end{equation}
where $\ket \psi$ is an uniform pure state on $\mcH \tens \mcK$ of distribution $\nu_{nk}$. 
\end{definition}

In order to get a better understanding of the measure $\mu_{n,k}$, we write $\psi$ in an orthonormal basis $\{e_i \tens f_j; 1 \leq i \leq n, 1 \leq j \leq k\}$ of $\mcH \tens \mcK$:
\begin{equation}
\psi = \sum_{i=1}^{n}{\sum_{j=1}^{k}{\psi_{ij}\, e_i \tens f_j}}.
\end{equation}
Thus the matrix $\ketbra{\psi}{\psi}$ has coordinates (in the same basis): 
\begin{equation}
\ketbra{\psi}{\psi}_{ij;i'j'}= \frac{\psi_{ij}\, \overline{\psi_{i'j'}}}{\sum_{\alpha=1}^{n}{\sum_{\beta=1}^{k}{\module{\psi_{\alpha\beta}}^2}}}.
\end{equation}
After taking the partial trace, we obtain
\begin{equation}
\rho_{ii'} = \frac{\sum_{j=1}^{k}{\psi_{ij}\, \overline{\psi_{i'j}}}}{\sum_{\alpha=1}^{n}{\sum_{\beta=1}^{k}{\module{\psi_{\alpha\beta}}^2}}}.
\end{equation}
Now, if we arrange the coordinates $\psi_{ij}$ of $\psi$ in a $n \times k$ matrix $X$ such that $X(i,j) = \psi_{ij}$, we have 
\begin{equation}
\rho = \frac{X\cdot X^*}{\trace(X\cdot X^*)}.
\end{equation}

Several important remarks should be made at this point. First of all, consider $U \in \mcU(n)$ and the density matrix $\rho'$ obtained by replacing $\psi$ with $(U\tens \I_\mcK)\psi$:
\begin{equation}
\rho' = \trace_\mcK(\ketbra{(U\tens \I_\mcK)\psi}{(U\tens \I_\mcK)\psi}).
\end{equation}
By the properties of the partial trace, we have that $\rho' = U \rho U^*$. But recall that the law of $\ket \psi$ is unitarily invariant; it is thus invariant by $U\tens \I_\mcK$ (which is an element of $\mcU(nk)$). Hence the law $\mu_{n,k}$ is invariant by unitary conjugation. Being positive, and thus self-adjoint, density matrices diagonalize:
\begin{equation}
\rho = V D V^*,
\end{equation}
with $V$ an unitary and $D$ a diagonal matrix with positive entries. The unitary invariance of $\mu_{n,k}$ corresponds to the fact that $V$ is distributed along the Haar measure on $\mcU(n)$. Remains, of course, the question of the distribution of $D$, the diagonal matrix of eigenvalues, which will be answered in Section \ref{sec:spectrum} (see Proposition \ref{prop:eig_rho}).

Another important question concerns the law of the matrix $X$. Recall that the coordinates of $X$ are those of $\psi$, rearranged in a $n \times k$ matrix. Since the pure state $\ket \psi$ is distributed along the uniform measure $\nu_{nk}$, we know, by the second point of Proposition \ref{prop:gaussian}, that we can take for $\psi$ a complex Gaussian vector in $\C^{nk}$. Thus, the elements of $X$ are independent, complex Gaussian random variables.
\begin{lemma}\label{lem:rdm_wishart}
Let $X$ be a $n \times k$ complex matrix such that the entries are
independent identically distributed (i.i.d.) $\mathcal N_\C(0, 1)$ random variables. Then, the matrix
\begin{equation}
\rho = \frac{X\cdot X^*}{\trace(X\cdot X^*)}
\end{equation}
has distribution $\mu_{n,k}$.
\end{lemma}
This lemma motivates the study of matrices of type $W = X\cdot X^*$, which will be taken up in the next section.

\section{Wishart matrices. Results at fixed size}\label{sec:wishart_fixed_size}

\subsection{The Wishart ensemble}
This section is devoted to introducing the Wishart ensemble of random matrices. Introduced in the 1930's to study covariance matrices in statistics, Wishart matrices have found many applications, both theoretical (random matrix theory) and practical: principal component analysis, engineering, etc. Let us start by recalling the definition of the Wishart ensemble:
\begin{definition}
Let $X$ be a $n \times k$ complex matrix such that the entries are
i.i.d. $\mathcal N_\C(0, 1)$ random variables. The $n \times n$
matrix $W = X \cdot X^*$ is called a \emph{Wishart random matrix} of
parameters $n$ and $k$.
\end{definition}

In virtue of Lemma \ref{lem:rdm_wishart}, there is a strong connection between the distribution of Wishart matrices and the random density matrices we study. More precisely, if $W$ is a Wishart matrix, then 
\begin{equation}
\rho = \frac{W}{\trace W}
\end{equation}
has distribution $\mu_{n,k}$.

We shall give a list of results on Wishart matrices that will be used later in the study of random density matrices. As the results are rather classical in random matrix theory, we will not supply proofs, but only references to the original papers. 

We start with a result on the eigenvalues of a Wishart matrix. Being of the form $W = X \cdot X^*$, Wishart matrices are positive and thus they admit $n$ non-negative eigenvalues $\lambda_1, \ldots, \lambda_n$. The next proposition provides the distribution of the random vector $(\lambda_1, \ldots, \lambda_n)$ (see \cite{mehta}).

\begin{proposition}
Let $W$ be a random $n \times n$ Wishart matrix with parameters $n$ and $k$. Then the distribution of the eigenvalues $(\lambda_1, \ldots, \lambda_n)$ has a density with respect to the Lebesgue measure on $\R^n_+$ which is given by
\begin{equation}
\Phi^{\mathcal W}_{n,k}(\lambda_1, \ldots , \lambda_n) = C_{n,
k}^{\mathcal W} \exp\left(-\sum_{i = 1}^{n}{\lambda_i}\right)
\prod_{i=1}^{n}{\lambda_i^{k-n}} \Delta(\lambda)^2,
\end{equation}
where $C_{n, k}^{\mathcal W}$ is the constant $\left[ \prod_{j =
0}^{n-1}{\Gamma(n+1-j)\Gamma(k-j)}\right]^{-1} $ and
\begin{equation}
\Delta(\lambda) = \prod_{1 \leq i < j \leq n}{(\lambda_i -
\lambda_j)}.
\end{equation}
\end{proposition}

When studying large random matrices, one important question is to
what resembles the spectrum of a random matrix in the limit $n
\rightarrow \infty$? In order to answer such a question, one
introduces the \emph{empirical spectral measure}
\begin{equation}
L_n(W) = \frac{1}{n} \sum_{i = 1}^{n}{\delta_{\lambda_i}},
\end{equation}
which is a random probability measure (it depends on $W$, which is
random). It turns out that, almost surely, the random measures
$L_n(W)$ converge to a deterministic probability measure, the
Marchenko-Pastur distribution.
\begin{definition}\label{def:MP}
For $c \in ]0, \infty[$, we denote by $\mu_c$ the
\emph{Marchenko-Pastur} probability measure given by the equation
\begin{equation}
\mu_c = \max\{1-c, 0\}\delta_0 + \frac{\sqrt{(x-a)(b-x)}}{2\pi
x}\indicator{[a,b]}(x)dx,
\end{equation}
where $a=(\sqrt c -1)^2$ and $b=(\sqrt c +1)^2$.
\end{definition}

The result is contained in the following theorem (see \cite{haag}).
\begin{theorem}\label{thm:Wishart_MP}
Assume that $c \in ]0, \infty[$, and let $(k(n))_n$ be a sequence of
integers such that $\lim_{n \rightarrow \infty}k(n)/n = c$. Consider
a sequence of random matrices $(W_n)_n$ such that for all $n$, $W_n$
is a Wishart matrix of parameters $n$ and $k(n)$. Define the
renormalized empirical eigenvalue distribution of $W_n$ by
\begin{equation*}
L_n = \frac{1}{n} \sum_{i = 1}^{n}{\delta_{n^{-1}\lambda_i(W_n)}},
\end{equation*}
where $\lambda_1(W_n), \cdots,\lambda_n(W_n)$ are the eigenvalues of
$W_n$. Then, almost surely, the sequence $(L_n)_n$ converges weakly
to the Marchenko-Pastur distribution $\mu_c$.
\end{theorem}

Another object of interest in random matrix theory is the largest eigenvalue of a large random matrix. The next result shows that in the Wishart case, it converges almost surely to the right edge of the support of the Marchenko-Pastur distribution; similarly to the Central Limit Theorem, the nature of the fluctuations is known (see \cite{bai} and \cite{johnstone}).

\begin{theorem}\label{thm:Wishart_largest}
Assume that $c \in ]0, \infty[$, and let $(k(n))_n$ be a sequence of integers such that $\lim_{n \rightarrow \infty}k(n)/n = c$. Consider a sequence of random matrices $(W_n)_n$ such that for all $n$, $W_n$ is a Wishart matrix of parameters $n$ and $k(n)$, and let $\lambda_{max}(W_n)$ be the largest eigenvalue of $W_n$. Then, almost surely,
\begin{equation}
\lim_{n \rightarrow \infty} \frac{1}{n}\lambda_{max}(W_n) = (\sqrt c +1)^2.
\end{equation}
Moreover, the following limit holds in distribution
\begin{equation}
\lim_{n \rightarrow \infty} \frac{\lambda_{max}(W_n) -n(\sqrt c
+1)^2}{n^{1/3}(1+\sqrt c)(1+1/\sqrt c)^{1/3}} = \mathcal W_2.
\end{equation}
\end{theorem}
Here, $\mathcal W_2$ is the Tracy-Widom law of order 2; as even the definition of this probability distribution is well beyond the scope of this work, we encourage the reader to look it up in \cite{tw}, the original paper of Tracy and Widom.

\subsection{The spectrum of a density matrix}\label{sec:spectrum}
Recall from Section \ref{sec:the_induced_measures} that when considering the diagonalization of a random density matrix
\begin{equation}
\rho = VDV^*,
\end{equation}
the unitary matrix $V$ is distributed along the Haar measure on the unitary group $\mcU(n)$. In this section we compute the distribution of the diagonal matrix $D$, i.e. the spectrum of a density matrix with distribution $\mu_{n,k}$. 

Here, as well as in the next section, the parameters $n$ and $k$ will be fixed, and we shall assume that $k \geq n$. If $n>k$, by a property of the partial trace application, the matrix $\rho$ will have $n-k$ null eigenvalues and $k$ eigenvalues identical to those of the density matrix
\[\sigma = \trace_\mcH(\ketbra\psi\psi).\]
In consequence, the study of the spectrum of $\rho$ is equivalent to the study of the spectrum of $\sigma$. Moreover, the size of $\sigma$'s environment ($n$) is larger than the dimension of $\sigma$ itself ($k$), and we can apply the first case. In conclusion, whenever $n$ is larger than $k$, we interchange $n$ and $k$, and we append $n-k$ null eigenvalues to the spectrum of $\rho$.

Recall that if $W$ is a Wishart matrix of parameters $n$ and $k$, then $\rho = W / \trace W$ has distribution $\mu_{n, k}$. It follows that if $(\lambda_1, \ldots, \lambda_n)$ are the eigenvalues of $W$ and $(\tilde \lambda_1, \ldots, \tilde \lambda_n)$ are those of $\rho$, then we have
\begin{equation}
\tilde \lambda_i = \frac{\lambda_i}{\sum_{j = 1}^{n}{\lambda_j}},
\quad \forall 1 \leq i \leq n.
\end{equation}

As the trace of a density matrix equals one, the (random) vector $(\tilde \lambda_1, \ldots, \tilde \lambda_n)$ is confined in the $(n-1)$-dimensional probability simplex $\Sigma_{n-1} = \{(x_1, \cdots, x_n) \in \R_+^n : \sum_{i=1}^{n}{x_i} = 1\}$. Note that $\tilde \lambda_n$ is a function of $\tilde \lambda_1, \ldots, \tilde \lambda_{n-1}$, so we will show that the distribution of $(\tilde\lambda_1, \ldots, \tilde\lambda_{n-1})$ admits a density w.r.t. the Lebesgue measure on $\Sigma_{n-1}$.

\begin{proposition}\label{prop:eig_rho}
The distribution of the (unordered) eigenvalues $\tilde\lambda_1(\rho), \ldots, \tilde\lambda_{n-1}(\rho)$ has a density with respect to the Lebesgue measure on $\Sigma_{n-1}$ given by
\begin{equation}\label{eqn:density_rho}
\Phi_{n,k}(\tilde \lambda_1, \ldots, \tilde \lambda_{n-1}) = C_{n,
k}   \prod_{i=1}^{n}{(\tilde \lambda_i)^{k-n}} \Delta(\tilde
\lambda)^2,
\end{equation}
where
\begin{equation}
C_{n, k} =  \frac{\Gamma(nk)}{\prod_{j =
0}^{n-1}{\Gamma(n+1-j)\Gamma(k-j)}}.
\end{equation}
\end{proposition}
\begin{remark}
In the formula (\ref{eqn:density_rho}), there are only $n-1$ variables; $\tilde \lambda_n$ is not a variable, but merely the notation $\tilde \lambda_n = 1 - (\tilde \lambda_1+ \cdots+ \tilde \lambda_{n-1})$.
\end{remark}
\begin{proof}
Let us start from the Wishart distribution of eigenvalues and consider the change of variables
\begin{align}
(\lambda_1, \ldots, \lambda_n) \mapsto (\lambda_1, \ldots, \lambda_{n-1}, S) \mapsto \\
\mapsto (\lambda_1/S, \ldots, \lambda_{n-1}/S, S) = (\tilde \lambda_1, \ldots, \tilde \lambda_{n-1}, S),
\end{align}
where $S=\sum_{i=1}^{n}{\lambda_i}$ is the sum of the Wishart eigenvalues. The Jacobian of this transformation equals $1/S^{n-1}$, and we get
\begin{equation}
\Phi^{(\tilde \lambda, S)}_{n,k}(\tilde \lambda_1, \ldots, \tilde
\lambda_{n-1}, S) = C_{n, k}^{\mathcal W} \exp(-S)
\prod_{i=1}^{n}{(S\tilde \lambda_i)^{k-n}} \Delta(S \tilde
\lambda)^2 S^{n-1}.
\end{equation}
We get now to the crucial point of the proof. Clearly, the above
density factorizes as
\begin{equation}\label{eqn:density_fact}
\Phi^{(\tilde \lambda, S)}_{n,k}(\tilde \lambda_1, \ldots, \tilde
\lambda_{n-1}, S) = C_{n, k}^{\mathcal W} \times \left[
\prod_{i=1}^{n}{\tilde \lambda_i^{k-n}} \Delta(\tilde
\lambda)^2\right]  \times \left[ S^{nk-1}\exp(-S)\right] .
\end{equation}
Hence, the normalized eigenvalues $(\tilde \lambda_1, \ldots, \tilde
\lambda_{n-1})$ and the sum of the Wishart eigenvalues $S$ are
\emph{independent} random variables.

In order to compute the distribution of $(\tilde \lambda_1, \ldots,
\tilde \lambda_{n-1})$, it suffices to take the marginal with
respect to $S$; we get
\begin{equation}
\Phi_{n,k}(\tilde \lambda_1, \ldots, \tilde \lambda_{n-1}) = C_{n,
k}   \prod_{i=1}^{n}{\tilde \lambda_i^{k-n}} \Delta(\tilde
\lambda)^2,
\end{equation}
where
\begin{align}
C_{n, k} &= C_{n, k}^{\mathcal W} \cdot \int_{0}^{\infty}{S^{nk-1} e^{-S} dS} = \Gamma(nk) C_{n, k}^{\mathcal W} =\\
&=\frac{\Gamma(nk)}{\prod_{j = 0}^{n-1}{\Gamma(n+1-j)\Gamma(k-j)}}.
\end{align}
\end{proof}

\begin{figure}[htb]
\centering
\includegraphics[width=3.9cm]{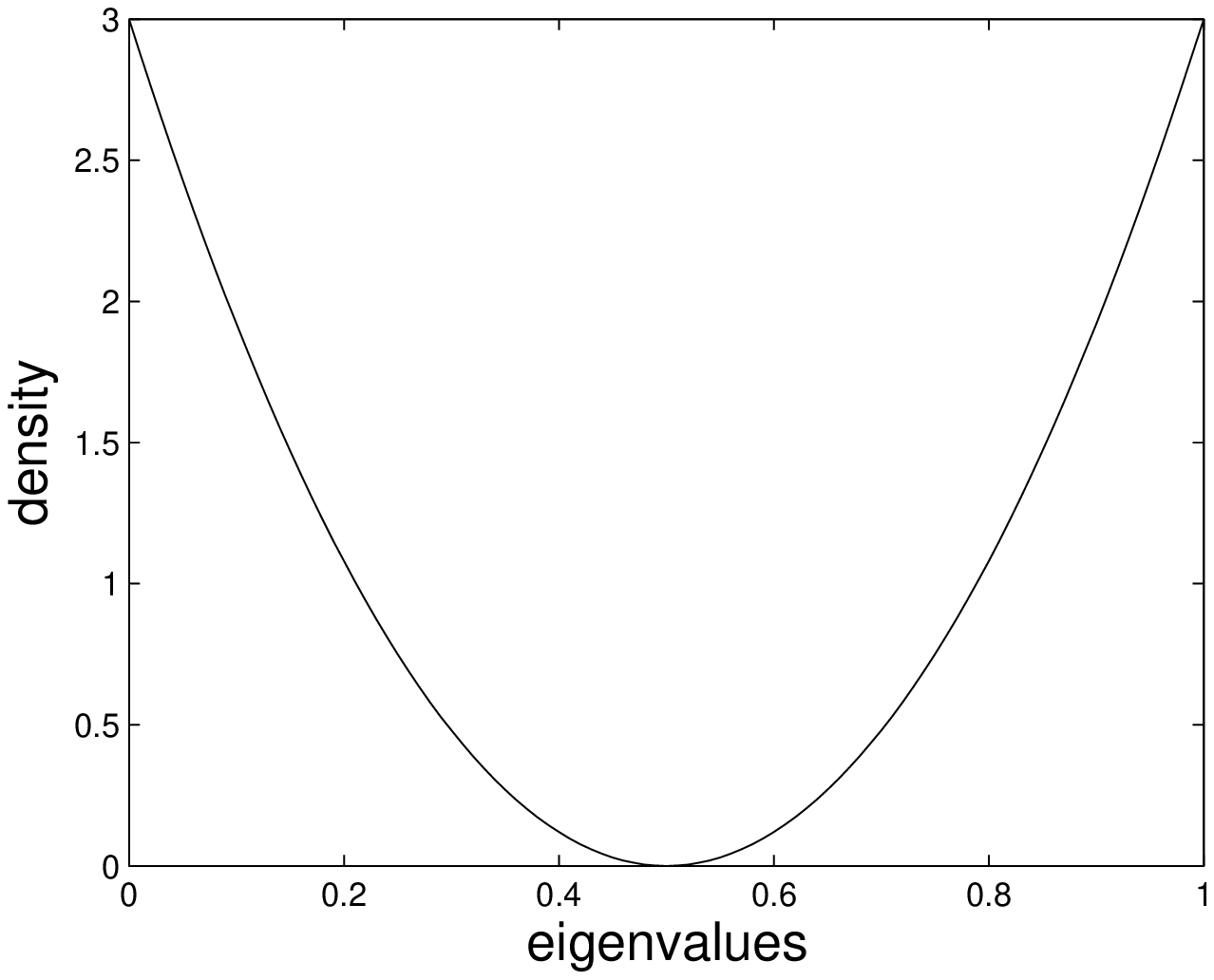}
\includegraphics[width=3.9cm]{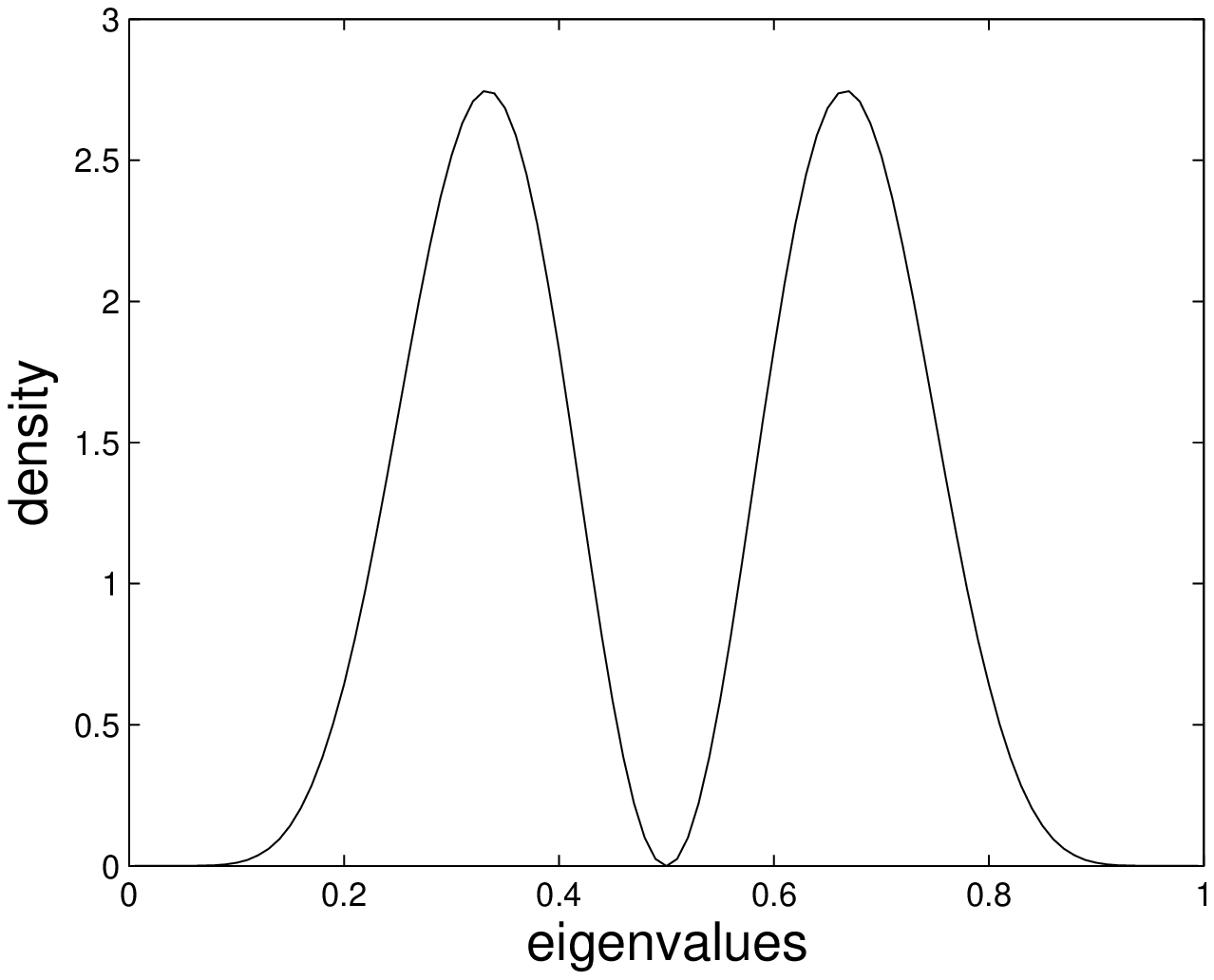}
\includegraphics[width=3.9cm]{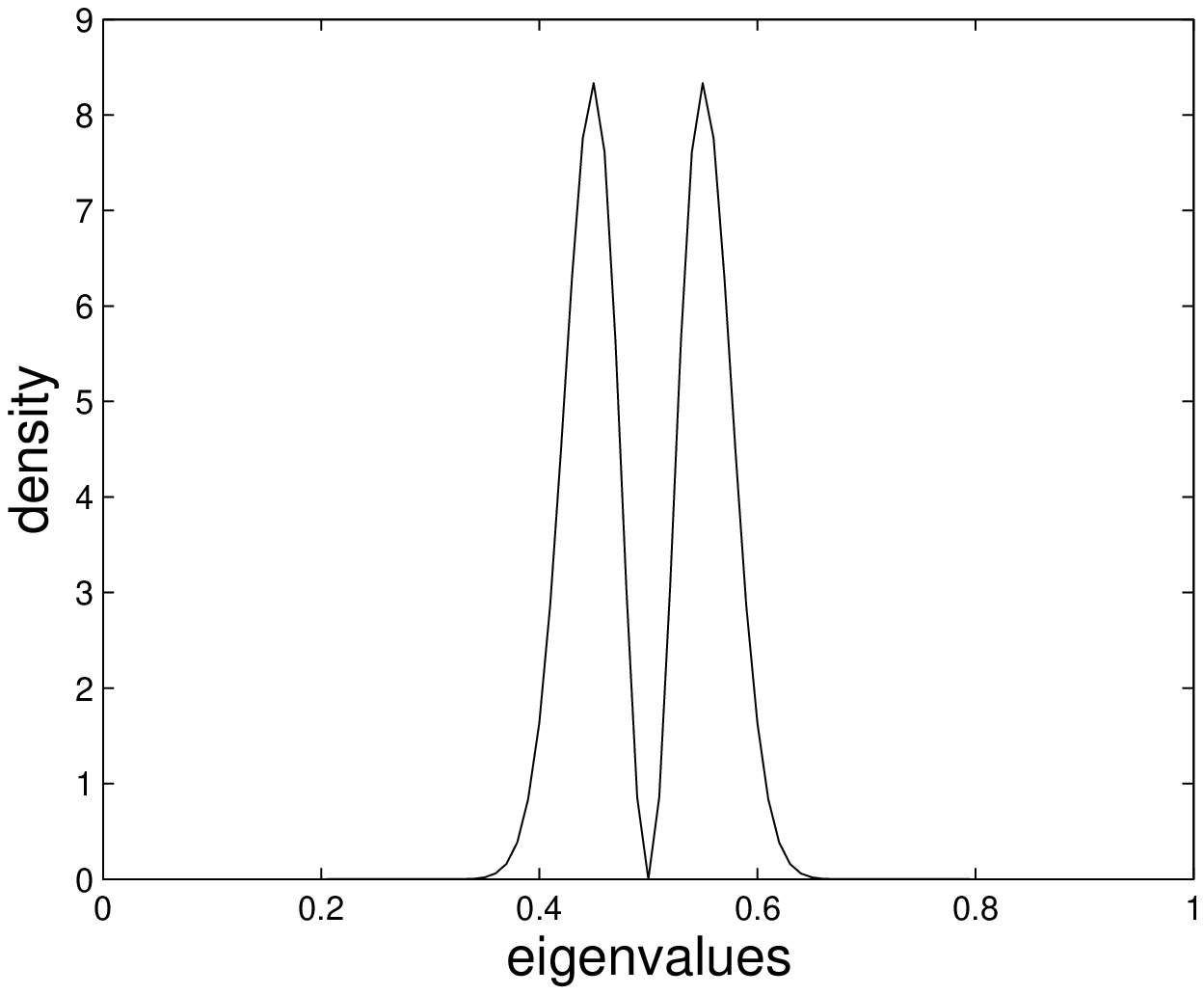}
\caption{Theoretical eigenvalue distribution for $(n=2,k=2)$, $(n=2,k=10)$ and $(n=2,k=50)$.}
\label{fig:n_2}
\end{figure}

\begin{figure}[htb]
\centering
\includegraphics[height=3.5cm]{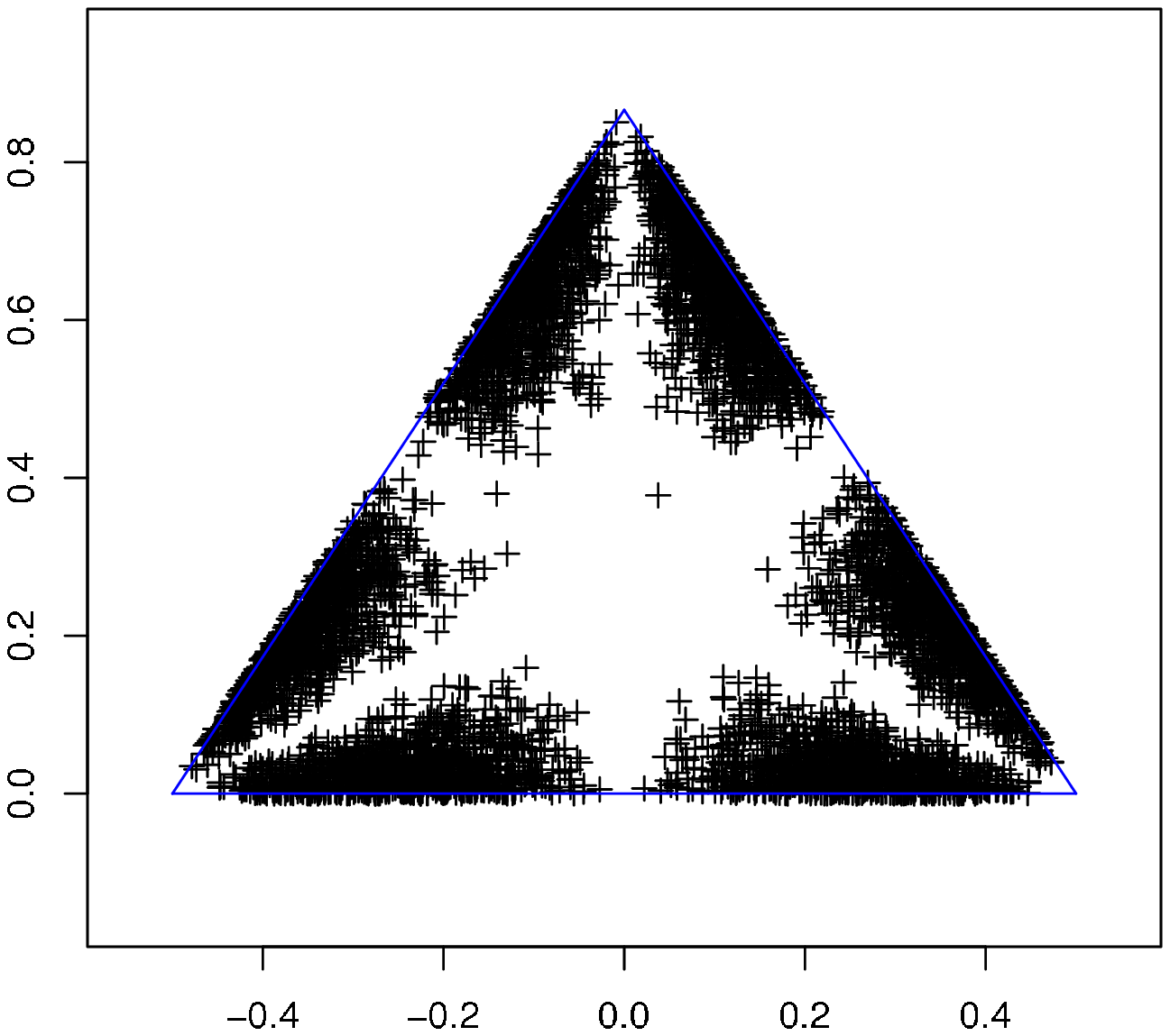}
\includegraphics[height=3.5cm]{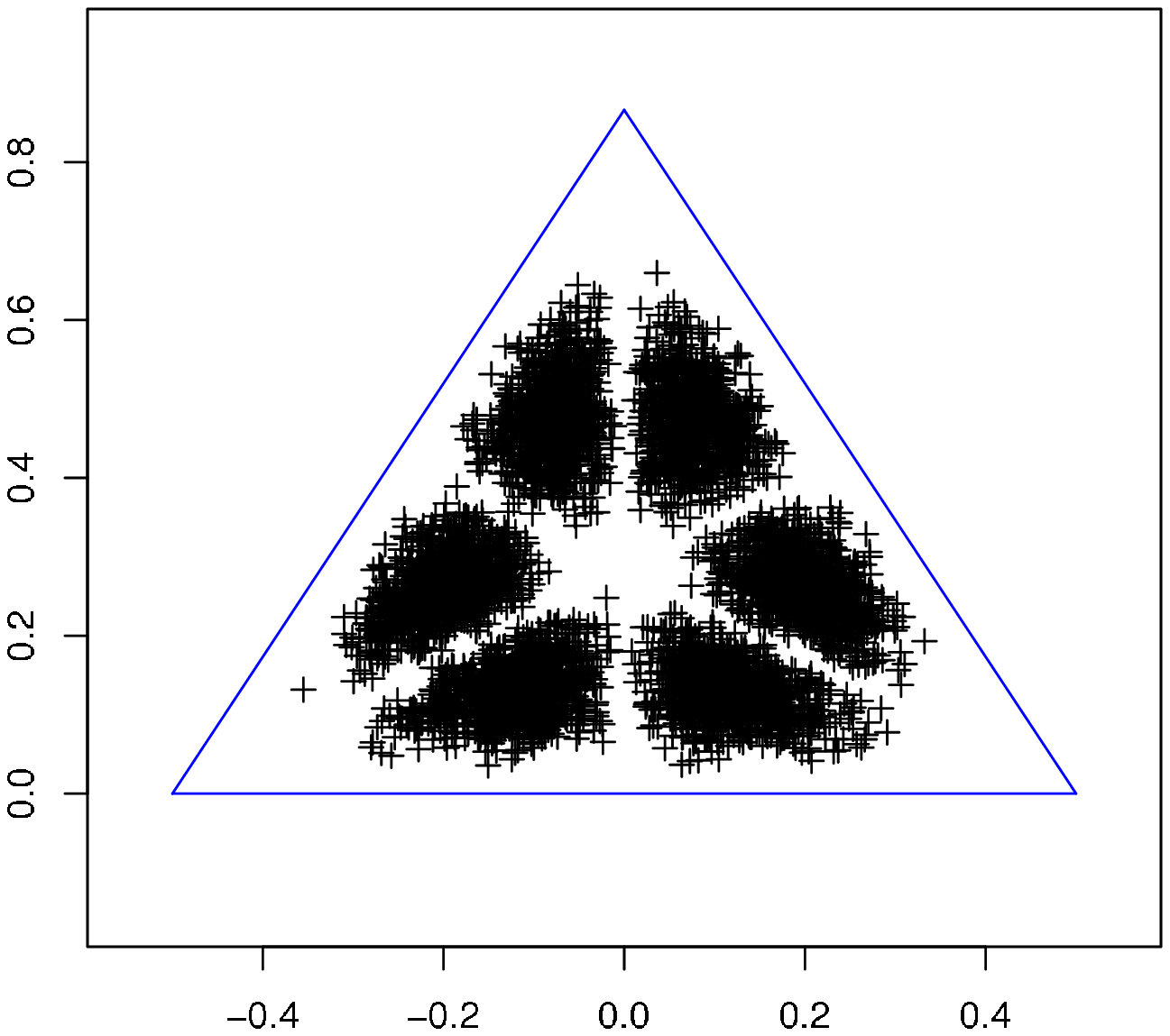}
\includegraphics[height=3.5cm]{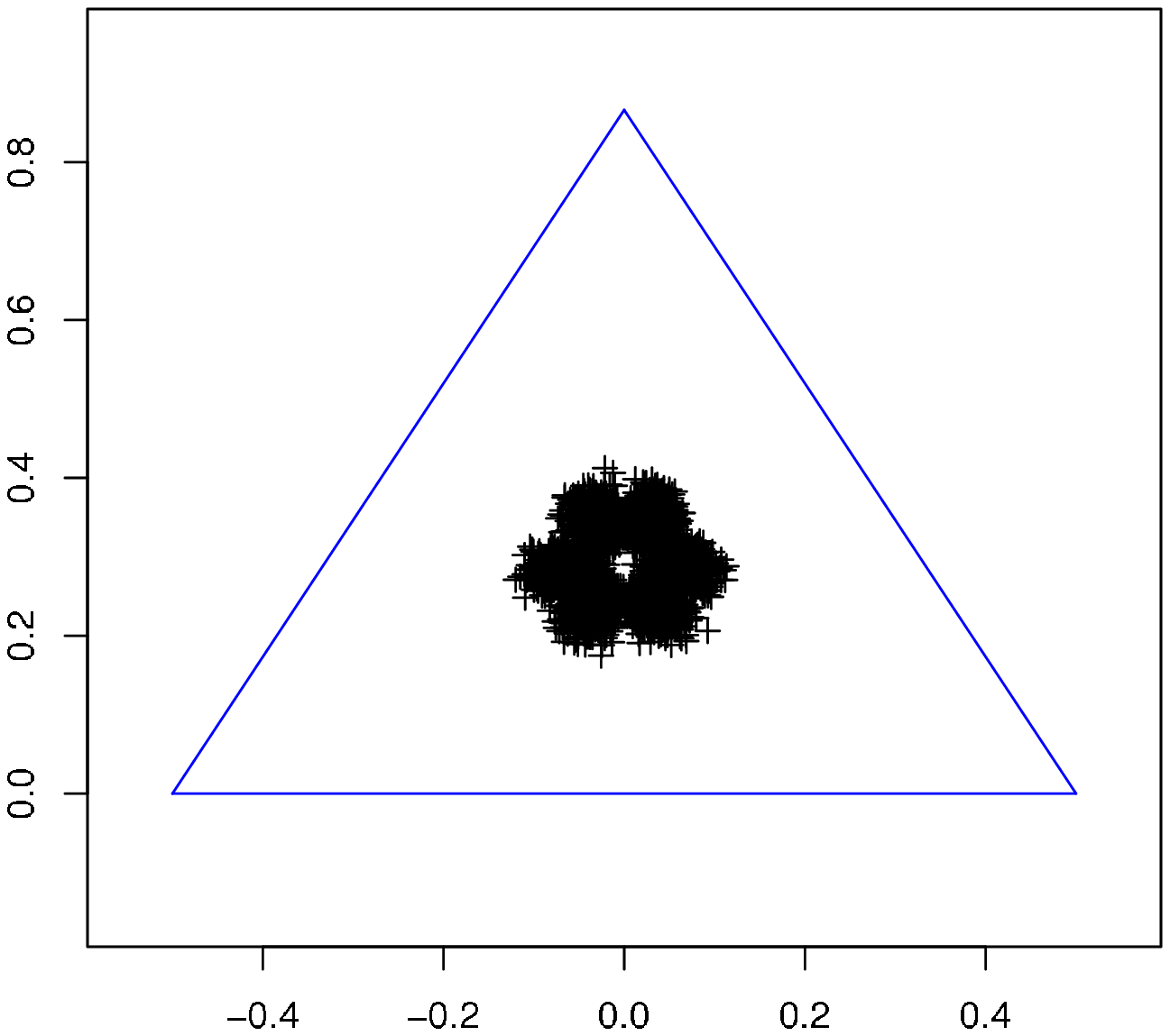}
\caption{Empirical eigenvalue distribution (5000 matrices) for $(n=3,k=3)$, $(n=3,k=10)$ and $(n=3,k=100)$.}
\label{fig:n_3}
\end{figure}

As a byproduct of the proof, we also obtain the following characterisation of the induced measure.
\begin{corollary}
The law of a random density matrix $\rho$ of parameters $n$ and $k$ is the law of a Wishart matrix $W$ of the same parameters \emph{conditioned} by $\trace W = 1$.
\end{corollary}
\begin{proof}
From the formula (\ref{eqn:density_fact}) we see that the normalized eigenvalues and the trace of a Wishart matrix are independent random variables. Thus, taking the marginal with respect to the trace is equivalent to conditioning on the event $(\trace W = 1)$. Note however that $(\trace W = 1)$ has zero probability.
\end{proof}
In the Figure \ref{fig:n_2} we have plotted the density functions for $n=2$ and several values of $k$ using the analytic formula (\ref{eqn:density_rho}). For $n=3$ we have randomly generated random density matrices and plotted the probability simplex $\Sigma_2$ along with the points corresponding to the spectra (Figure \ref{fig:n_3}). We notice that for large values of $k$ (the size of the environment), the spectrum concentrates to the middle point in $\Sigma_{n-1}$. This is a general phenomenon and it will be studied in section \ref{sec:first_model}.

\subsection{Moments} 
The aim of this section is to provide formulas for the moments of order $q$ of a random density matrix of distribution $\mu_{n,k}$. In order to do that, we shall introduce the some notation: $\E_{n,k}[\cdot]$ will denote the expectation with respect to the law $\mu_{n,k}$ and $\E_{n,k}^{\mathcal W}[\cdot]$ the expectation with respect to the law of Wishart matrices with parameters $n$ and $k$. We will use the corresponding result on the Wishart ensemble and derive explicit formulas, as well as recurrence relations. The following proposition provides a bridge between the moments of a density matrix and those of a Wishart matrix with the same parameters.

\begin{proposition}
Let $\E_{n,k}[\trace(\rho^q)]$ be the moment of a random density
matrix of parameters $n$ and $k$ and let $\E_{n,k}^{\mathcal
W}[\trace(W^q)]$ be the moment of a Wishart matrix having the same
parameters. Then,
\begin{equation}\label{eqn:moments}
\E_{n,k}[\trace (\rho^q)] = \frac{\E_{n,k}^{\mathcal W}[\trace
(W^q)]}{nk(nk+1)\cdots(nk+q-1)}.
\end{equation}
\end{proposition}
\begin{proof}
By using the same technique as in the proof of the Proposition
\ref{prop:eig_rho}, we get
\begin{equation}
\E_{n,k}^{\mathcal W}[\trace (W^q)] = \E_{n,k}[\trace (\rho^q)]
\frac{\Gamma(nk+q)}{\Gamma(nk)},
\end{equation}
which is the same as equation (\ref{eqn:moments}).
\end{proof}

We can find in the literature different explicit and recurrence
formulas for $\E_{n,k}^{\mathcal W}[\trace (W^q)]$. From the one in
\cite{haag}, we get
\begin{equation}\label{eqn:exact_moments}
\E_{n,k}[\trace (\rho^q)] =
\frac{\Gamma(nk)}{\Gamma(nk+q)}\sum_{j=1}^{q}{(-1)^{j-1}
\frac{[k+q-j]_q [n+q-j]_q }{(q-j)! (j-1)!}},
\end{equation}
where $[a]_q = a(a-1)\cdots(a-q+1)$. The recurrence formula (see \cite{haag})
\begin{align}\label{eqn:rec_rel}
\E_{n,k}[\trace (\rho^{q})] &=
\frac{(2q-1)(n+k)}{(nk+q-1)(q+1)}\E_{n,k}[\trace (\rho^{q-1})] +\\
&+\frac{(q-2)((q-1)^2-(k-n)^2)}{(nk+q-1)(nk+q-2)(q+1)}\E_{n,k}[\trace
(\rho^{q-2})]
\end{align}
allows us to easily compute some averages:
\begin{align}
&\E_{n,k}[\trace (\rho^2)] = \frac{n+k}{nk+1},\\
&\E_{n,k}[\trace(\rho^3)] = \frac{n^2+3nk+k^2+1}{(nk+1)(nk+2)},\\
&\E_{n,k}[\trace (\rho^4)] = \frac{n^3+6n^2k+6nk^2+k^3+5n+5k}{(nk+1)(nk+2)(nk+3)}, \quad etc.
\end{align}

These formulas are consistent with the ones of \cite{sz1} and \cite{sz2}.

\section{Asymptotics}\label{sec:asymptotics}
The last part of this paper is devoted to the study of random
density matrices corresponding to \emph{large systems}. We shall
consider two models, both motivated physically:

\begin{enumerate}
\item In the first model, the size of the density matrix $n$ is constant and the size of the environment $k$ tends to infinity. Such a situation arises typically when one studies a small system (a qubit, a pair of qubits, etc.) coupled to a much larger environment. We show that in the limit $k \rightarrow \infty$, density matrices distributed along $\mu_{n, k}$ converge to the maximally mixed (or chaotic) state $\I/n$. 
\item In the second model, both $n$ and $k$ tend to infinity and $k/n \rightarrow c>0$. This model describes a large system coupled to a large environment with constant ratio of size ($\dim \mathcal K / \dim \mcH \approx c$). In this case we show that the spectral measures of density matrices of law $\mu_{n,k}$ converge to a deterministic measure known in random matrix theory as the \emph{Marchenko-Pastur distribution} (see Definition \ref{def:MP}). We also study the convergence and the fluctuations of the largest eigenvalue of random density matrices.
\end{enumerate}

\subsection{The first model}\label{sec:first_model}
Consider the density function of $\mu_{n,k}$ with $n$ fixed and $k
\rightarrow \infty$:
\begin{equation}
\Phi_{n,k}(\lambda_1, \ldots, \lambda_{n-1}) = C_{n, k}
\prod_{i=1}^{n}{\lambda_i^{k-n}} \Delta(\lambda)^2.
\end{equation}
As $n$  is fixed, the Vandermonde factor $\Delta(\lambda)$ is
constant. The other factor, properly normalized in order to get a
probability density, is the Dirichlet measure of parameter $\alpha =
k-n+1$:
\begin{equation}
\Phi_{n,k}^{'}(\lambda_1, \ldots, \lambda_{n-1}) =
C^{'}_{n,k}\prod_{i=1}^{n}{\lambda_i^{\alpha-1}}.
\end{equation}

The next result is well-known in probability theory. We shall sketch its proof for the sake of completenss.
\begin{theorem}
The Dirichlet measure converges weakly as $\alpha \rightarrow
\infty$ to the Dirac measure $\delta_{(1/n, \ldots, 1/n)}$
\end{theorem}
\begin{proof}
The idea behind the proof is to show that the variance of a Dirichlet-distributed random variable converges to 0 as its parameter converges to infinity. Let $X$ be such a random variable. $X$ has a density with respect to the Lebesgue measure on the probability simplex given by:
\[f(x_1, \ldots, x_n) = \frac{\Gamma(n\alpha)}{\Gamma(\alpha)^n} \prod_{i=1}^{n}{x_i^{\alpha-1}}.\]
It is easy to compute
\begin{align}
\E\left[\norm{X-\left(\frac 1 n, \ldots, \frac 1 n\right)}^2\right] &= n\E\left[x_1^2 - \frac{2x_1}{n} + \frac{1}{n^2}  \right] =  \frac{\alpha + 1}{n\alpha +1} - \frac{1}{n} \rightarrow 0.
\end{align}
\end{proof}

As the maximally mixed state $\I/n$ is the unique state having
spectrum $\{1/n, \ldots, 1/n\}$, we get:
\begin{corollary}
Density matrices of the first model converge almost surely to the
maximally mixed (or chaotic) state $\I/n$.
\end{corollary}
\begin{remark}
The same result can be obtained by an entropic argument. It turns out that the mean von Neumann entropy $S(\rho) = - \trace(\rho \log \rho)$ can be computed for a random density matrix distributed along $\mu_{n,k}$:
\[\E_{n,k}[S(\rho)] = \sum_{i = k+1}^{nk}{\frac 1 i} - \frac{n-1}{2k}.\]
This formula has been conjectured by Page \cite{page} and has been subsequently proved (see \cite{sanchez, sen}) using various methods. Let us explain how it implies the corollary. First, fix $n$ and let $k$ grow to infinity, as in our model. The mean entropy is easily seen to converge to $\log n$. This turns out to be the maximum von Neumann entropy for a system with $n$ degrees of freedom. It is attained at the state $I/n$, the unique state of maximum uncertainty.
\end{remark}
\subsection{The second model}
In the second model, both the size of the density matrix and the size of the environment tend to infinity. In order to use the results on the Wishart ensemble (Theorems \ref{thm:Wishart_MP} and \ref{thm:Wishart_largest}), we need appropriate results on the behavior of the trace $S$ of a Wishart matrix.
\begin{lemma}\label{lem:Wishart_trace}
Assume that $c \in ]0, \infty[$, and let $(k(n))_n$ be a sequence of
integers such that $\lim_{n \rightarrow \infty}k(n)/n = c$. Consider
a sequence of random matrices $(W_n)_n$ such that for all $n$, $W_n$
is a Wishart matrix of parameters $n$ and $k(n)$. Let $S_n = \trace
W_n$ be the trace of $W_n$. Then
\begin{equation}
\frac{S_n}{nk(n)} \rightarrow 1  \quad \text{almost surely}
\end{equation}
and
\begin{equation}
\frac{S_n - nk(n)}{\sqrt{nk(n)}} \Rightarrow \mathcal N(0,1),
\end{equation}
where $``\Rightarrow"$ denotes the convergence in distribution.
\end{lemma}

\begin{proof}
Recall that $W_n = X_n \cdot X_n^*$, when $X_n$ is a $n \times k(n)$
matrix with i.i.d. complex Gaussian entries. We have
\begin{equation}
S_n = \sum_{i=1}^{n}\sum_{j=1}^{k(n)} \module{X_{ij}}^2 =
\sum_{i=1}^{n}\sum_{j=1}^{k(n)} (\Re(X_{ij})^2 +\Im(X_{ij})^2).
\end{equation}
The random variables $\{\Re(X_{ij}), \Im(X_{ij})\}_{ij}$ are i.i.d.
with distribution $\mathcal N(0, 1/2)$ and thus, by the law of large
numbers, we have, almost surely,
\begin{equation}
\lim_{n \rightarrow \infty} \frac{S_n}{2nk(n)} = \frac{1}{2},
\end{equation}
completing the proof of the first result. The second result follows
from the Central Limit Theorem.
\end{proof}

We can now state and prove the analogue of Theorem
\ref{thm:Wishart_MP} for random density matrices:

\begin{theorem}\label{thm:density_MP}
Assume that $c \in ]0, \infty[$, and let $(k(n))_n$ be a sequence of
integers such that $\lim_{n \rightarrow \infty}k(n)/n = c$. Consider
a sequence of random density matrices $(\rho_n)_n$ such that for all
$n$, $\rho_n$ has distribution $\mu_{n, k(n)}$. Define the
renormalized empirical distribution of $\rho_n$ by
\begin{equation}
L_n = \frac{1}{n} \sum_{i = 1}^{n}{\delta_{cn\lambda_i(\rho_n)}},
\end{equation}
where $\lambda_1(\rho_n), \cdots ,\lambda_n(\rho_n)$ are the
eigenvalues of $\rho_n$. Then, almost surely, the sequence $(L_n)_n$
converges weakly to the Marchenko-Pastur distribution $\mu_c$.
\end{theorem}
\begin{proof}
We know (Theorem \ref{thm:Wishart_MP}) that the empirical
distribution of eigenvalues for the Wishart ensemble
\begin{equation}\label{eqn:sp_measure_Wishart}
L_n^{\mathcal W} = \frac{1}{n} \sum_{i =
1}^{n}{\delta_{n^{-1}\lambda_i(W_n)}},
\end{equation}
converges almost surely to the Marchenko-Pastur distribution of
parameter $c$. Recall that the eigenvalues of the density matrix
$\rho_n = W_n/ \trace(W_n)$ are those of $W_n$ divided by the trace
$S_n$ of $W_n$. We have thus the following formula for the empirical
spectral measure of $\rho$:
\begin{equation}
L_n = \frac{1}{n} \sum_{i = 1}^{n}{\delta_{cn\lambda_i(W_n)/S_n}} =
\frac{1}{n} \sum_{i = 1}^{n}{\delta_{n^{-1}\lambda_i(W_n) \cdot
\frac{cn^2}{S_n}}}.
\end{equation}
The last equation is the same as equation (\ref{eqn:sp_measure_Wishart}) with the Dirac measures perturbed by a factor of $cn^2/S_n$ which converges, almost surely, to 1 (by the preceding lemma). We are now going to show that such a perturbation does not change the limit in distribution. In order to achieve this, recall that when the limit measure is compactly supported, the convergence in distribution is equivalent to the convergence of moments. If we compute the $q$-th moment of the measures $L_n^{\mathcal W}$ and $L_n$, we find:
\begin{equation}
\langle x^q,L_n^{\mathcal W} \rangle = \frac{1}{n}\sum_{i=1}^{n}{\left(n^{-1}\lambda_i(W_n)\right)^q},
\end{equation}
and, respectively,
\begin{equation}
\langle x^q,L_n \rangle = \frac{1}{n}\sum_{i=1}^{n}{\left(n^{-1}\lambda_i(W_n)\right)^q}\cdot \left(\frac{cn^2}{S_n}\right)^q.
\end{equation}
These expressions have the same limit as $n \rightarrow \infty$ for
all $q$, and thus $L_n$ converges to the Marchenko-Pastur
distribution.
\end{proof}

\begin{figure}[htb]
\centering
\includegraphics[width=3.9cm]{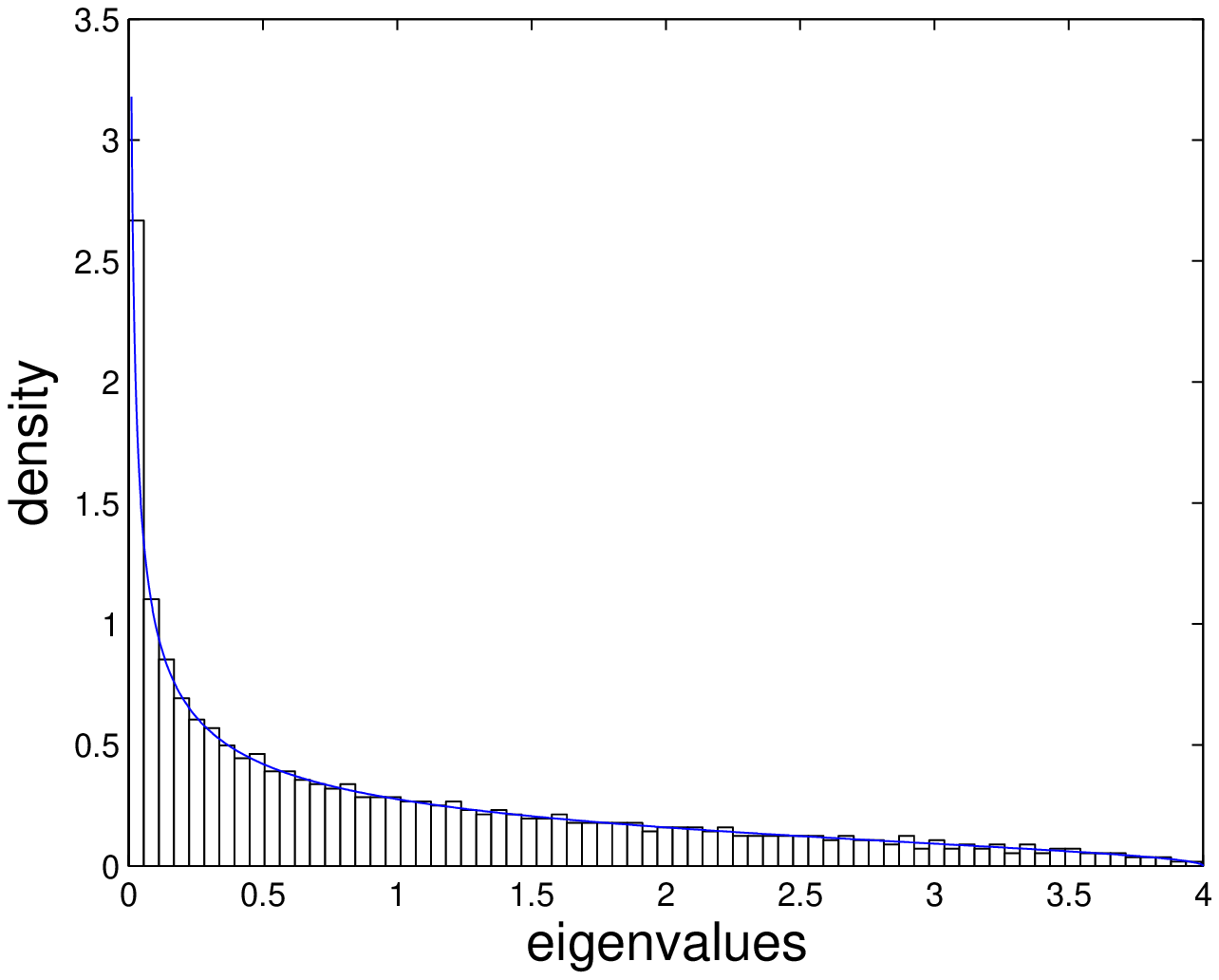}
\includegraphics[width=3.9cm]{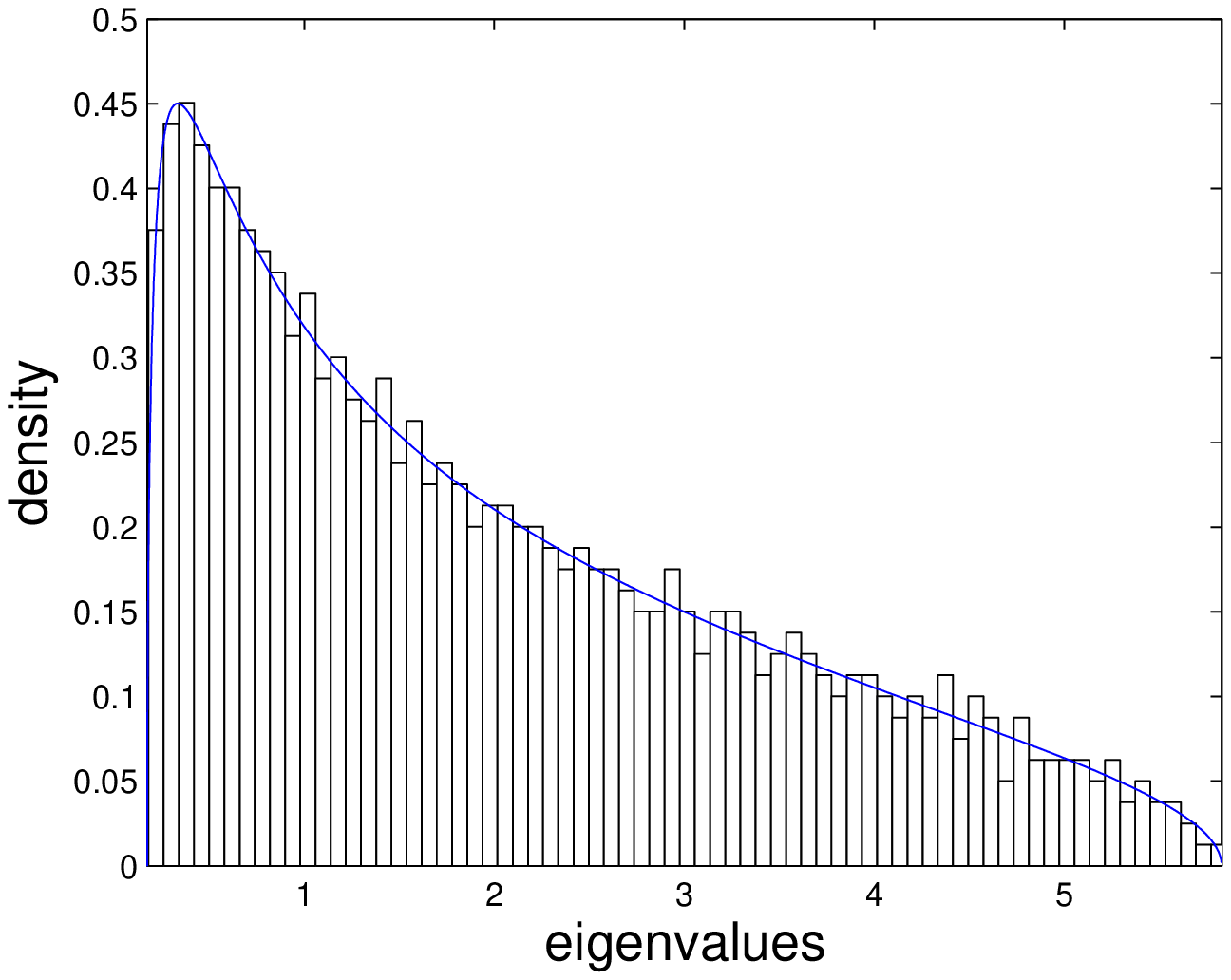}
\includegraphics[width=3.9cm]{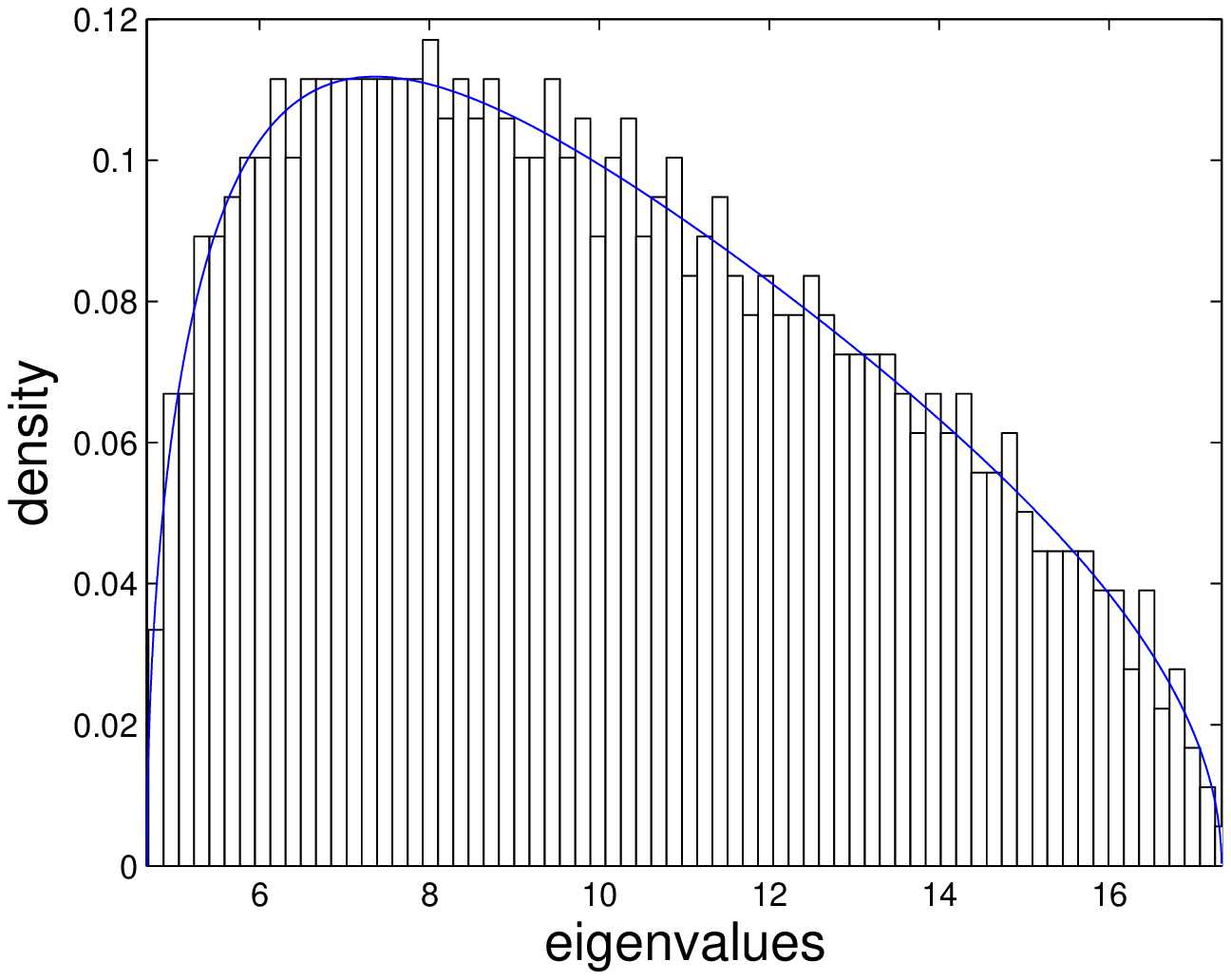}
\caption{Empirical and limit measures for $(n=1000,k=1000)$, $(n=1000,k=2000)$ and $(n=1000,k=10000)$.}
\label{fig:MP}
\end{figure}

In the Figure \ref{fig:MP}, we have plotted for several values of $c$ and large $n$ and $k$ a histogram of the spectrum for \emph{one} density matrix and the theoretical density of the Marchenko-Pastur distribution (see Remark \ref{rk:almost_sure}). We can see that the empirical histogram matches closely the theoretical curve for rather mild values of $n$ (here $n=1000$).

We now turn to the study of the largest eigenvalue of random density
matrices. As before, we use the corresponding result on the Wishart ensemble (Theorem \ref{thm:Wishart_largest}) and the control over the trace (Lemma \ref{lem:Wishart_trace}):

\begin{theorem}\label{thm:density_largest}
Assume that $c \in ]0, \infty[$, and let $(k(n))_n$ be a sequence of integers such that $\lim_{n \rightarrow \infty} k(n)/n = c$. Consider a sequence of random matrices $(\rho_n)_n$ such that for all $n$, $\rho_n$ has distribution $\mu_{n,k(n)}$, and let $\lambda_{max}(\rho_n)$ be the largest eigenvalue of $\rho_n$. Then, almost surely, 
\begin{equation}\label{eqn:largest_eigen_rho}
\lim_{n \rightarrow \infty} cn\lambda_{max}(\rho_n) = (\sqrt c +1)^2.
\end{equation}
Moreover,
\begin{equation}\label{eqn:fluct_largest}
\lim_{n \rightarrow \infty} \frac{n^{2/3}\left[cn\lambda_{max}(\rho_n) -(\sqrt c+1)^2\right]}{(1+\sqrt c)(1+1/\sqrt c)^{1/3}} = \mathcal W_2 \quad \text {in distribution}.
\end{equation}
\end{theorem}
\begin{proof}
By the first part of Theorem \ref{thm:Wishart_largest}, the (normalized) largest eigenvalue $\frac{1}{n}\lambda_{max}(W_n)$ of a Wishart matrix converges almost surely to $(\sqrt c +1)^2$. Obviously, we have
\begin{equation}
\lambda_{max}(\rho_n) = \frac{\lambda_{max}(W_n)}{S_n},
\end{equation}
and, by the Lemma \ref{lem:Wishart_trace}, $S_n / (cn^2)$ converges (almost surely) to 1. Finally, we obtain formula (\ref{eqn:largest_eigen_rho}). 

For the second part of the theorem, what we need to do, normalizations apart, is to show that the trace of a Wishart matrix fluctuates less than the largest eigenvalue. For the Wishart case, we have
\begin{equation}
\lambda_{max}(W_n) = n(\sqrt c +1)^2 + n^{1/3}(1+\sqrt c)(1+1/\sqrt c)^{1/3}(\mathcal W_2 + \littleo(1)),
\end{equation}
and
\begin{equation}
S_n =  nk(n) + \sqrt{nk(n)}(\mathcal N + \littleo(1)).
\end{equation}
Again, $\lambda_{max}(\rho_n) = \lambda_{max}(W_n) / S_n$ and after simplifications, one obtains the desired formula (\ref{eqn:fluct_largest}).
\end{proof}
\begin{remark}\label{rk:almost_sure}
Note that Theorem \ref{thm:density_MP} and the first part of Theorem \ref{thm:density_largest} deal with \emph{almost sure} convergences. This means that when considering sequences of random density matrices of increasing size, the respective convergences will fail only on a set of null measure. This is to be compared with typicality results for random density matrices obtained recently in \cite{lebowitz}, \cite{popescu} by concentration of measure techniques. These results give bounds (at fixed matrix size) on the probability that a random matrix is far from its expected value, while our results deal with the more subtle convergence of rescaled quantities, such as the spectral distribution or the largest eigenvalue.
\end{remark}

\section{Conclusions}
We investigated random density matrices distributed along the so-called \emph{induced measures}. After introducing them as partial traces of larger random pure states, we provided some explicit and recurrence relations for the moments of such density matrices. Using results on Wishart matrices, we then considered large density matrices. In a first model, a fixed size system was coupled to a very large environment; we showed that an uniform pure state on the compound system corresponds to the maximally mixed (or chaotic) density matrix on the fixed-size system. In parallel with Wishart matrices, we studied the regime $\dim \mcK / \dim \mcH \rightarrow c$. We obtained the almost sure convergence of the empirical spectral measure and of the largest eigenvalue, as well as the fluctuations of the largest eigenvalue. Results from random matrix theory were easily adapted for density matrices. Other important quantities, such as correlation functions, require a more detailed analysis, and  this will be the subject of further work. Also, it may be interesting to study such asymptotics for other probability measures on density matrices, such as the Bures measure.

\textbf{Acknowledgment:} The author would like to thank Guillaume Aubrun for useful ideas which led to several simplifications in some proofs.

\end{document}